\documentclass[% 
reprint,
amsmath,
amssymb,
aps,
prb,
floatfix,
]{revtex4-2}

\usepackage[english]{babel}
\usepackage{amsmath}  % needed for \tfrac, \bmatrix, etc.
\usepackage{amsfonts} % needed for bold Greek, Fraktur, and blackboard bold
\usepackage{graphicx} % needed for figures
\usepackage{hyperref}
\usepackage{natbib}
\usepackage{array}
\usepackage[table]{xcolor}
\usepackage{ulem}
\usepackage{xcolor}
\usepackage{braket}
\usepackage{breqn}
\usepackage{upgreek}
\usepackage{caption}

\begin{document}

\captionsetup[figure]{name={FIG.},labelsep=period,justification=raggedright,font=small}
\captionsetup[table]{name={TABLE},labelsep=period,justification=raggedright,font=small}

\title{Quantum Computation Protocol for Dressed Spins in a Global Field}

\author{Amanda E. Seedhouse} 
\author{Ingvild Hansen} 
\author{Arne Laucht} 
\author{Chih Hwan Yang} 
\author{Andrew S. Dzurak}
\author{Andre Saraiva} 
\affiliation{School of Electrical Engineering and Telecommunications$,$ The University of New South Wales$,$ Sydney$,$ NSW 2052$,$ Australia}

\begin{abstract}
Spin qubits are contenders for scalable quantum computation because of their long coherence times demonstrated in a variety of materials, but individual control by frequency-selective addressing using pulsed spin resonance creates severe technical challenges for scaling up to many qubits. This individual resonance control strategy requires each spin to have a distinguishable frequency, imposing a maximum number of spins that can be individually driven before qubit crosstalk becomes unavoidable. Here we describe a complete strategy for controlling a large array of spins in quantum dots dressed by an on-resonance global field, namely a field that is \textit{constantly driving} the spin qubits, to dynamically decouple from the effects of background magnetic field fluctuations. This approach -- previously implemented for the control of single electron spins bound to electrons in impurities -- is here harmonized with all other operations necessary for universal quantum computing with spins in quantum dots.
We define the logical states as the dressed qubit states and discuss initialization and readout utilizing Pauli spin blockade, as well as single- and two-qubit control in the new basis. Finally, we critically analyze the limitations imposed by qubit variability and potential strategies to improve performance.
\end{abstract}

 
\maketitle{} 

\section{Introduction}
Electron spins in semiconductors are a quintessential example of desirable qubits, with long coherence times and excellent controllability based on either magnetic or electric dipole spin resonance with microwaves~\cite{pla2012single,kawakami2014electrical, koppens2006driven, laird2013valley, nadj2010spin, watzinger2018germanium,maurand2016cmos}. In devices with a few qubits, individual spins can be addressed by selectively driving their unique transition frequencies. However, this approach is limited because the spread of resonance frequencies is determined by material characteristics, which sets a maximum range and therefore a maximal number of spins one can drive before being plagued by qubit crosstalk. The issue of frequency crowding has been studied for transmon qubits~\cite{Schutjens2013single} and can be alleviated in spin qubits using engineered narrow-band microwave pulses. With an increase in number of spin qubits, however, this would become increasingly difficult and require long and complex engineered pulses.

Pursuing full scale quantum computing has its challenges. There have been many proposals for scaling up different types of qubits~\cite{loss_quantum_1998, fowler_surface_2012, veldhorst_silicon_2017,  hill_surface_2015, lekitsch_blueprint_2017, li_crossbar_2018, kane_silicon-based_1998}, with some~\cite{veldhorst_silicon_2017, kane_silicon-based_1998, hill_surface_2015, lekitsch_blueprint_2017, li_crossbar_2018} that recognize the potential of using an off-resonance global field. This field may, for example, be generated by a three-dimensional cavity resonator that subjects the whole chip to the oscillatory magnetic field~\cite{kane_silicon-based_1998, Vahapoglu2021}. The aim of the off-resonance global field is that qubits defined in the idle state will be spins that are not resonant with the field. To perform qubit operations the spins are brought into resonance with the global field. This can be done in different ways depending on the type of qubit and architecture, for example by controlling the electrical Stark shift for tuning of the qubit resonance frequency \cite{laucht_2015_electrically}. 

Here, we explore in detail the use of an \textit{on-resonance} global field, in which qubits are considered idle when they are being constantly driven by the global field, called dressed qubit states \cite{ Mollow1969PowerSystems,Xu2007CoherentDot,Baur2009MeasurementQubit,London2013DetectingSpin,laucht_dressed_2017,Huang2021Electric-dipole-inducedDot}. Microwave dressing of qubit states increases coherence times due to the constant decoupling from background environmental fluctuations in the system \cite{laucht_dressed_2017}. Specifically, we analyze the qubit dressing strategies available for spins in semiconductor quantum dots and harmonize all the other operations (initialization, readout and two-qubit gates) with the always-on field. We study single-qubit operations discussing two strategies that we call frequency-shift keying and frequency modulation. Two-qubit gates leveraging the Heisenberg spin-spin interaction are also explored, showing the evolution from the SWAP gate to the CPHASE gate as the system parameters are changed. Throughout the whole quantum computation, the dressing field is always on, meaning the protocol should include initialization and readout in the dressed basis. Limitations of each process are discussed, including how variability in qubit frequencies can affect the globally driven system.

\section{Overview of spin array architectures}
\begin{figure*}
    \centering
    \includegraphics[width=2\columnwidth]{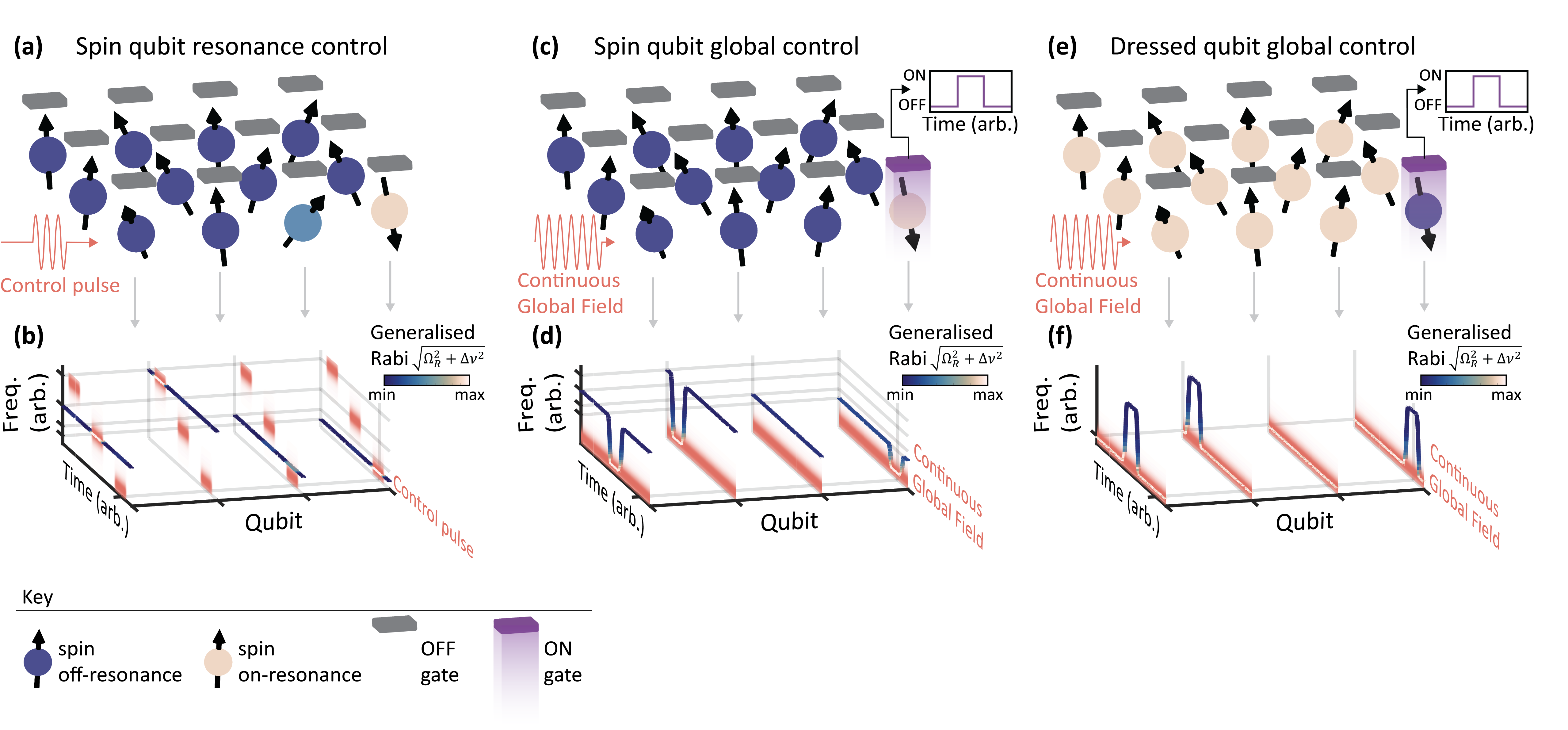}
    \caption{Comparing three methods of controlling spin qubits in the context of scalable quantum systems. (a) A two-dimensional array of spin qubits with microwave-pulsed (pink arrow) single qubit control mechanism. A key describing the array of qubits is given at the bottom of the figure. The light blue spin demonstrates unintentional interactions with the microwave field. (b) The four closest spins have their frequencies plotted as a function of time where the color map of the line shows the generalized Rabi frequency resulting from the microwave control pulse. Three ESR control pulses (marked in pink) are applied to three spins, the marked line on the time axis indicates the snippet of time that (a) is showing. For a global microwave control field (c) a two-dimensional array of qubits is shown, now with the qubit control achieved via switching the voltage bias on a gate electrode to enable a Stark shift, shown by the purple square pulse. The frequencies of four qubits (d) with the frequency (Stark) shift switched on when qubit addressing takes place. On-resonance dressed spins (e) with control mediated by pulsing a gate with four tracked frequencies (f), where the qubit addressing is achieved by a gate-voltage-enabled Stark shift to take the spin out of resonance with the global microwave field.}
    \label{fig:global_scalable}
\end{figure*}

Selective control of individual spin qubits becomes limiting if the control mechanism uses pulsed microwave electron spin resonance (ESR) and addresses each spin by their unique Larmor frequency. The Larmor frequency of a spin in a semiconductor device is determined by the microscopic environment that surrounds the spin. For example, in the case of electron spin qubits in purified silicon $^{28}$Si, in which the reduced presence of nuclear spins leads to small Overhauser fields, the spin-orbit interaction is the leading mechanism that results in the spread of qubit frequencies~\cite{tanttu_controlling_2019, Ruskov2018ElectronDots, Ferdous2018Interface-inducedScalability}. In GaAs, on the other hand, the Overhauser field has an important role in setting the spin qubit frequency~\cite{bluhm2011dephasing}. For ESR with frequency selection to work in a large array of spins, each qubit frequency should be separated by several times the Rabi frequency in order to avoid errors that can degrade the quality of fault tolerant operations. Some strategies may remedy these errors through the use of pulse shaping~\cite{yang_silicon_2019}, or by engineering the difference in Larmor frequencies with a magnetic field gradient introduced by a micromagnet~\cite{Pioro-Ladriere2008ElectricallyField,Wu2014Two-axisMicromagnet}, but ultimately there will be difficulty in individual addressing for scaled up systems. A more scalable pathway, as discussed next, is to locally control the spin-orbit coupling by applying electric fields with gate electrodes in order to dynamically control the value of the spin resonance frequency.

For scalable quantum computing, we need arrays of qubits such as the one schematically shown in Fig.~\ref{fig:global_scalable}(a). Each spin may be either off-resonance or on-resonance with a electromagnetic driving wave, which will mean that it is either simply precessing (off-resonance) or nutating resonantly (on-resonance), in which case the spin may rotate between $\ket{\uparrow}$ and $\ket{\downarrow}$. The dynamical state of each spin (driven or undriven) is represented by the colour of a sphere, where dark blue shows an undriven spin and beige represents a driven spin. These colours represent the generalised Rabi frequency for off-resonant driving $\Omega_{\rm R}^{\rm gen}(\Omega_{\rm R},\Delta\nu)=\sqrt{\Omega_{\rm R}^2+\Delta\nu^2}$, where $\Omega_{\rm R}$ is the Rabi frequency and $\Delta\nu$ the detuning of the qubit Larmor frequency from the microwave frequency. The cuboids represent gate electrodes, with grey representing those that are used only for forming quantum dots but are not activating any targeted shift in the spin resonance. Purple means the gate has been biased to a state that purposely shifts the spin frequency of an electron by locally controlling its spin-orbit coupling. The pink arrow shows a microwave control pulse that is targeting control on a single qubit, but simultaneously acts on all qubits of the array. The diagram shows the qubit array for a single time instance during the control operation. The target qubit is the beige qubit, showing resonance control. 

The main difficulty in this strategy for a spin qubit architecture is frequency crowding -- the statistical dispersion of the qubit frequencies is limited, such that eventually the separation between two qubit frequencies become smaller than the ESR linewidth. The light-blue spin has a Larmor frequency close to the target qubit resulting in unwanted rotations, corresponding to qubit crosstalk. This is more clear in Fig~\ref{fig:global_scalable}(b) which shows a full time trace as a function of frequency of four spins, with each spin frequency represented by a solid line, and the arrows show which trace corresponds to which qubit. We represent the condition for a spin to be considered on resonance as a color map that shows the amplitude of $\Omega_{\rm R}^{\rm gen}$ if we assume that the initial state was either $\ket{\uparrow}$ or $\ket{\downarrow}$. This amplitude is maximum when the microwave pulse frequency (marked in pink) matches the qubit frequency, and decays as we detune the spin in a range set by the Rabi frequency. The control pulses of the two leftmost qubits are performed with no crosstalk, but the two rightmost qubits are similar in frequency so that when one qubit is targeted, there is significant off-resonant driving on the neighbouring resonance. The line indicated on the time axis shows the time instance that Fig~\ref{fig:global_scalable}(a) represents.

To improve upon this method, the idea of an off-resonance global field has been studied in the literature~\cite{veldhorst_silicon_2017, kane_silicon-based_1998, hill_surface_2015, lekitsch_blueprint_2017, li_crossbar_2018}. There, the global field is always on and the qubits are considered to be in the idle state when they are out of resonance with the magnetic field. To perform qubit operations, the spins are individually brought into resonance with the field by some method that locally controls the qubit frequency. This can be performed, for example, by electrically controlling the frequency shift caused by the hyperfine or spin-orbit interactions. Using a gate electrode~\cite{kane_silicon-based_1998, Jones2018LogicalDots}, each spin frequency can be addressed by locally controlling the $g$-factor~\cite{li_crossbar_2018}, the overlap between the electron wavefunction with a nuclear spin~\cite{Hensen2020AQubit}, or a combination of the two~\cite{laucht_2015_electrically}.

Single qubit control using an off-resonance global field is shown schematically in Fig.~\ref{fig:global_scalable}(c). In the figure, the electric control is represented by a gate electrode (grey square) that can be switched ON (purple) using a voltage pulse to change the Stark shift of a single qubit, thus changing its $g$-factor. The change in $g$-factor is chosen to bring the qubit into resonance with the global field, allowing for rotations to occur. This method removes the issue of having individual resonance frequencies for each qubit, reducing crosstalk effects (represented in the figure where only a single qubit is being driven). Figure.~\ref{fig:global_scalable}(d) shows three rotations performed on qubits in an off-resonance global field. When each of the rotations are performed, the individual qubits are brought into resonance, matching the always-on global field frequency. 

This method has some limitations. Firstly, it relies on the ability to have all spins out of resonance with the global field initially, and to be brought back to resonance on demand. The first condition is achieved by guaranteeing that the microwave frequency $f_{\rm mw}$ and the frequency of each of the individual spins $\nu$ is separated by well more than the Rabi frequency, $f_{\rm mw}-\nu\gg\Omega_R$. This alone can be easily achieved by a proper choice of external magnetic field $B_0$, which controls the distribution of values of $\nu$. However, the electric controllability of the individual $\nu$ needs then to be sufficient to bring each of the spins back into resonance. Electrons at a Si/SiO$_2$ interface, for instance, have a typical Stark shift $d\nu/dV$ in the range of $\pm$10MHz/V for magnetic fields near $B_0=$1T. The range of applicable voltage pulses is typically in the hundreds of millivolts, set by the quantum dot charge and orbital transitions, and is preferably kept to a minimum to avoid disturbances in the electrostatic landscape caused by agitating charged defects. Only the qubits with the largest Stark shift will, therefore, have enough range to be operated in this manner.

This leads us to the method proposed in this paper: an on-resonance global field where the qubits are tuned so that they are constantly being driven. This creates dressed qubits~\cite{Mollow1969PowerSystems,Xu2007CoherentDot,Baur2009MeasurementQubit,London2013DetectingSpin,laucht_dressed_2017} which are defined in terms of the combination of the spin and the modes of the electromagnetic field. In the rotating frame, the driving field creates an energy splitting between the superposition states $\ket{z_\rho} = (1/\sqrt{2}) (\ket{\downarrow}+\ket{\uparrow})$ and $\ket{\bar{z}_\rho}  = (1/\sqrt{2}) (\ket{\downarrow}-\ket{\uparrow})$, and we use these to define the logical states $\ket{z_\rho}=\ket{0}$ and $\ket{\bar{z}_\rho}=\ket{1}$. The dressed states are represented by the beige coloured spins in Fig.~\ref{fig:global_scalable}(e). In Fig.~\ref{fig:global_scalable}(f) we represent one method by which single qubit operations could be performed in three different spins with an on-resonance global field. To perform qubit control a similar technique is used as for the off-resonance global control technique; the $g$-factor is shifted to remove the qubit from resonance, which allows for nutation between the dressed states~\cite{laucht_dressed_2017}. This approach combines the scalable addressability of the global field with the added advantage of decoupling from non-Markovian noise. Moreover, we will demonstrate that the range of control for the qubit frequencies $\nu$ can be made less stringent using this approach.

\section{Definition of the qubit}

In this section, the derivation of the dressed spin Hamiltonian is shown for the case of single- and two-qubit interactions. The basic concept of dressing with a single spin has been discussed in the literature~\cite{laucht_dressed_2017}, so we will be brief on this aspect. 

Considering a single spin qubit with a static magnetic field $B_0$ along $z$ and driving magnetic field $B_1$ along $x$, the Hamiltonian is as follows 
\begin{equation}
    H_{\textrm{lab}} =  \frac{g \mu_{\text{B}}}{2} [B_{0} \sigma_{z} + B_1 \cos{(2\pi f_{\text{mw}}t)} \sigma_{x}],
\end{equation}
where $\mu_{\text{B}}$ is the Bohr magneton, $g$ the electron spin $g$-factor, $f_{\text{mw}}$ the driving magnetic field frequency, $t$ the time, and $\sigma_j$ the Pauli matrices, where $j= x, y, z$. Following this, we move to the familiar rotating frame representation in which the frame rotates with the angular velocity of the microwave field, not the qubits,
\begin{equation}
    H_{\text{rot}} = \frac{h}{2} (\Delta\nu \sigma_{z} + \Omega_{\text{R}} \sigma_{x}),
    \label{eq:ham_2rot}
\end{equation}
where $h$ is the Planck constant, $\Delta\nu$ the detuning from the spin Larmor frequency $\nu$ ($\Delta\nu = g \mu_{\text{B}} B_0/h - f_{\text{mw}}= \nu - f_{\text{mw}}$), and $\Omega_{\text{R}}$ the Rabi frequency. It should be noted that throughout this text the frequencies denoted $f$ are that of instruments, $\nu$ the Larmor frequencies, and $\Omega_{\text{R}}$ the Rabi frequencies.

\begin{figure}
    \centering
    \includegraphics[width=1\columnwidth]{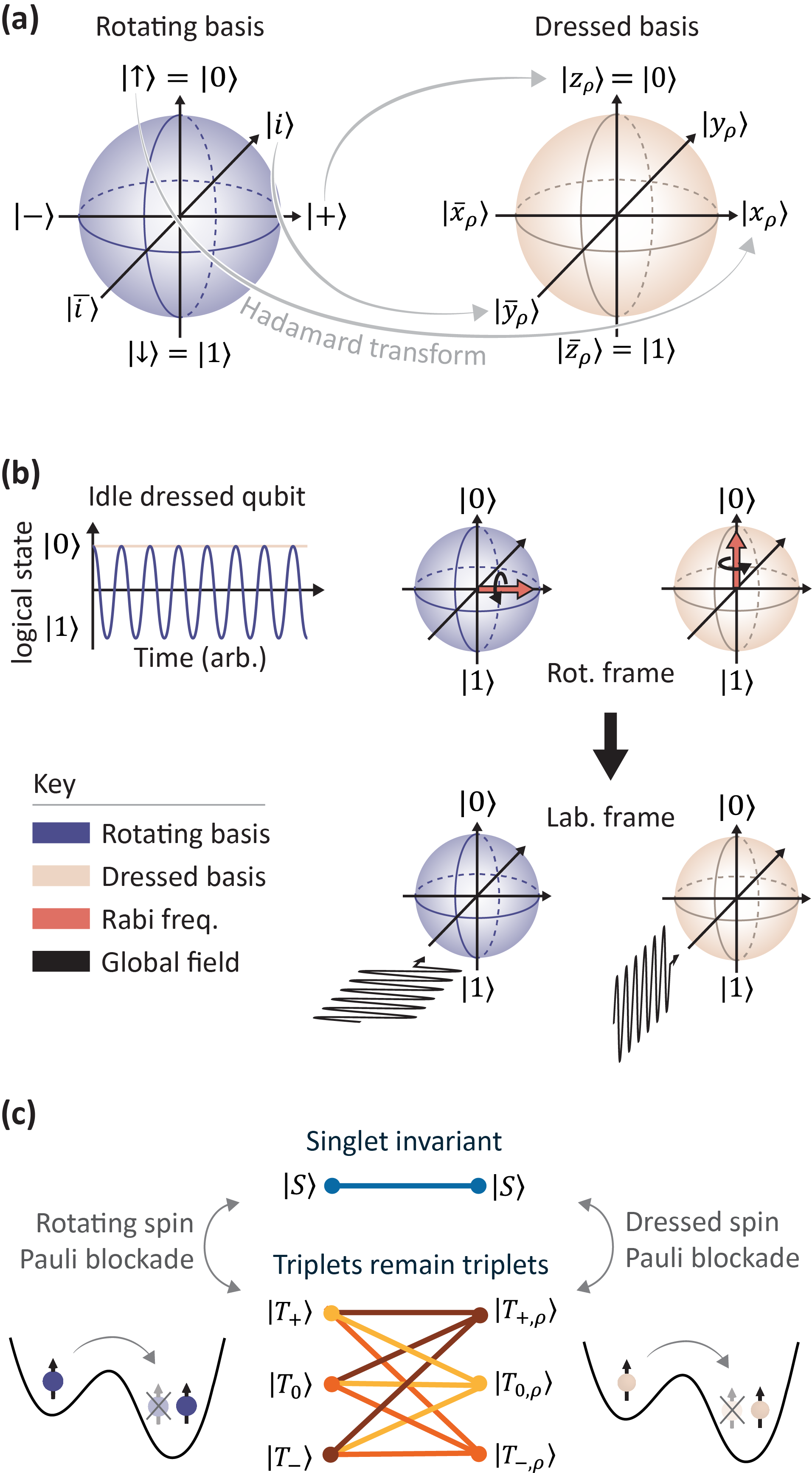}
    \caption{The process of dressing spin qubits. The Bloch sphere representation of a bare spin qubit in the rotating frame (blue) and a dressed spin (beige). The gray arrows show the transformation from bare to dressed via the Hadamard transformation. This transformation changes the two-qubit singlet and triplet states (b), showing $\ket{S}$ is invariant under the transformation and the triplet states remaining in the triplet family. This means that the Pauli spin blockade remains. The blockade in a double quantum dot system is demonstrated for (blue) a rotating frame bare spin triplet $\ket{\text{T}_+}$, and (beige) a dressed spin triplet $\ket{\text{T}_{+,\rho}}$.}
    \label{fig:dressing}
\end{figure}

The rotating frame Hamiltonian describes the traditional spin qubit with the quantisation axis along $\ket{\uparrow}$. A Bloch sphere representation of this is shown in Fig.~\ref{fig:dressing}(a) by the blue sphere, where the logical states, $\ket{0}$ and $\ket{1}$ are represented by $\ket{\uparrow}$ and $\ket{\downarrow}$, respectively. On the equator are the superposition states $\ket{+} = (1/\sqrt{2}) (\ket{\downarrow}+\ket{\uparrow})$, $\ket{-} = (1/\sqrt{2}) (\ket{\downarrow}-\ket{\uparrow})$, $\ket{i} = (1/\sqrt{2}) (\ket{\downarrow}+i\ket{\uparrow})$ and $\ket{\bar{i}} = (1/\sqrt{2}) (\ket{\downarrow}-i\ket{\uparrow})$. 

\begin{table}
    \centering
     \setlength{\tabcolsep}{0pt}
        \setlength\extrarowheight{2pt}
        \begin{tabular}{cc|cc}
        \hline
        \hline
        \multicolumn{2}{c|}{Rotating basis} & \multicolumn{2}{c}{Dressed basis}  \\
        \hline
        \multicolumn{1}{c}{$\hspace{0.1cm}$ Bloch axis $\hspace{0.3cm}$} &\multicolumn{1}{c|}{State$\hspace{0.1cm}$}   &\multicolumn{1}{c}{$\hspace{0.1cm}$ Bloch axis $\hspace{0.3cm}$}  & \multicolumn{1}{c}{State $\hspace{0.1cm}$}\\
        \hline
        $x$ & $\ket{+}$  & $z$ & $\ket{z_\rho}$\\
        $y$ & $\ket{i}$  & $-y$ & $\ket{\bar{y}_\rho}$ \\
        $z$ & $\ket{\uparrow}$  & $x$ &  $\ket{x_\rho}$ \\
        \hline
        \hline
    \end{tabular}
    \caption{The Bloch axis direction and states of the rotating frame and dressed frame states. Each row is equivalent.}
    \label{tab:tab}
\end{table}

To transform into the dressed picture, a constant amplitude $B_1$ field is applied. 
As discussed before, this creates an energy separation between $\ket{+}$ and $\ket{-}$ states, which in the rotating frame sets a new quantisation axis along $x$. To find the dressed basis Hamiltonian, $H_{\text{rot}}$ is transformed using the Hadamard unitary $U_{\rm (Hadamard)}$~\cite{nielsen2002quantum}. Writing $H_{\rho} = U_{\rm (Hadamard)} H_{\text{rot}}U_{\rm (Hadamard)}^\dagger$, we get
\begin{equation}
    H_{\rho} = \frac{h}{2} (\Omega_{\text{R}} \sigma_{z} +  \Delta\nu\sigma_{x}).
    \label{eq:single_dressed_qubit}
\end{equation}
The bare spin qubit $H_{\text{rot}}$ is transformed into the dressed spin qubit $H_{\rho}$, as shown in the Bloch sphere in Figs.~\ref{fig:dressing}(a) with the grey arrows. Table~\ref{tab:tab} shows the Bloch sphere axes and qubit state representation in the rotating basis and dressed basis, introducing $\ket{x_\rho}, \ket{y_\rho}, \ket{z_\rho}$. The logical states are now encoded as $\ket{z_\rho}=\ket{0}$ and $\ket{\bar{z}_\rho}=\ket{1}$. The dressed Hamiltonian $H_{\rho}$ tells us that the energy difference between the logical states is determined by the Rabi frequency $\Omega_{\text{R}}$, and the detuning of the driving field from the qubit frequency $\Delta\nu$ determines the qubit rotations.

It is important to understand that, in the dressed basis, if the spin is pointing along the equator it will precess due to the Rabi frequency. Figure~\ref{fig:dressing}(b) shows a plot of the logical state evolving in both the rotating bare spin and dressed spin basis with the global field on-resonance. The Bloch spheres to the right of the plot show the direction of the Rabi frequency amplitude in the rotating frame, as well as the laboratory frame $B_1$ magnetic field oscillations for both the rotating bare spin and dressed cases. 

To describe a universal gate set for dressed spins, the Hilbert space is expanded to include two-qubit interactions. Here, we denote the singlet states with an S and the triplet states with a T. When referring to dressed triplet states, the state will be followed by $\rho$ (this is not done for singlets because they are invariant under the dressing transformation). 

The system we will describe involves a double quantum dot with the charge occupation in the left and right dots indicated by $(N_1, N_2)$ where $N_1$ and $N_2$ are integers. In general, we are interested in either the situation where there are two electrons in the quantum dots or when the number of electrons is such that they form closed shells, with only two active spins not being inert. For simplicity, we omit the number of electrons in closed shells. If both active electrons are occupying the same quantum dot, we assume them to form a singlet state in the lowest orbital, following the Pauli exclusion principle, and that state is denoted $\ket{\text{S}(0,2)}$. For the (1,1) charge configuration, considering the rotating bare spin case, the other four levels are $\ket{\uparrow \uparrow},\ket{\uparrow \downarrow},\ket{\downarrow \uparrow},\ket{\downarrow \downarrow}$. 

The magnetic field vector in quantum dots (assumed to be the same) is represented in the lab frame as $\vec{B} = [B_1 \cos{(2\pi f_{\text{mw}}t)},0, B_0]$, and the Pauli matrices acting on the left (right) dot are given as $\vec{\sigma}_{1(2)}$, and the $g$-factor in the left (right) dot is $g_{1 (2)}$. The Hamiltonian describing the (1,1) occupied states in the rotating frame is
\begin{equation}
    H_{(1,1), \textrm{rot}} = \frac{h}{2} \big( \Delta\nu_1 \sigma_{z1}+ \Delta\nu_2 \sigma_{z2} + \Omega_{\textrm{R}1} \sigma_{x1} + \Omega_{\textrm{R}2} \sigma_{x2} \big),
\end{equation}
where the detuning from the Larmor frequency of qubit 1(2) is denoted $\Delta\nu_{1(2)}$ and the Rabi frequency of qubit 1(2) $\Omega_{\text{R}1(2)}$. The Pauli matrix acting on qubit 1(2) is $\sigma_{j1(2)}$. Then, the Hamiltonian can be transformed into the dressed basis $\{ \ket{\text{T}_{+,\rho}},\ket{z_\rho \bar{z}_\rho},\ket{\bar{z}_\rho z_\rho},\ket{\text{T}_{-,\rho}} \}$ using the transformation Hadamard $\otimes$ Hadamard,
\begin{equation}
    H_{(1,1), \rho} = \frac{h}{2} \big( \Omega_{\textrm{R}1} \sigma_{z1}+ \Omega_{\textrm{R}2} \sigma_{z2} + \Delta\nu_1 \sigma_{x1} + \Delta\nu_2 \sigma_{x2} \big).
\end{equation}
Singlet states are rotationally invariant, allowing the Pauli spin blockade to be preserved in the dressed, basis which is useful for initialization, readout and exchange interactions. Figure \ref{fig:dressing}(c) illustrates the blockade of $\ket{\text{T}_+}$ in the rotating bare spin picture (blue) and $\ket{\text{T}_{+,\rho}}$ (beige) in the dressed picture, and how the singlets remain invariant, while the triplets rearrange within the triplet manifold, shown with the colored lines.

In the dressed picture, we have the usual singlet subspace $\{\ket{\text{S}(1,1)},\ket{\text{S}(0,2)}\}$,
\begin{equation}
    H_{(0,2)\rightarrow(1,1)} =\frac{h}{2} 
    \begin{pmatrix}
    0 && 2t_{\text{c}}\\
    2t_{\text{c}} && -2\varepsilon 
    \end{pmatrix}.
\end{equation}
The tunnel coupling between the dots is given by $t_{\text{c}}$ and the chemical potential bias between the two quantum dots $\varepsilon$. Combining the singlet interactions with the basis $\{ \ket{\text{T}_{+,\rho}},\ket{z_\rho \bar{z}_\rho},\ket{\bar{z}_\rho z_\rho},\ket{\text{T}_{-,\rho}} \}$ gives the final 5 by 5 Hamiltonian in the dressed basis $\{\ket{\text{S}(0,2)}, \ket{\text{T}_{+,\rho}},\ket{z_\rho \bar{z}_\rho},\ket{\bar{z}_\rho z_\rho},\ket{\text{T}_{-,\rho}} \}$
\begin{widetext}
\begin{equation}
H_\rho =\frac{h}{2} 
    \begin{pmatrix}
     -2 \varepsilon && 0 && \sqrt{2} t_{\text{c}} && -\sqrt{2} t_{\text{c}} && 0\\
     0  && \Omega_{\textrm{R}1}+\Omega_{\textrm{R}2} && \Delta\nu_2 && \Delta\nu_1 && 0\\
     \sqrt{2}t_{\text{c}} && \Delta\nu_2 && \Omega_{\textrm{R}1}-\Omega_{\textrm{R}2} && 0 && \Delta\nu_1\\
     -\sqrt{2}t_{\text{c}} && \Delta\nu_1 && 0 && -\Omega_{\textrm{R}1}+\Omega_{\textrm{R}2} && \Delta\nu_2\\
     0 && 0 && \Delta\nu_1 && \Delta\nu_2 && -\Omega_{\textrm{R}1}-\Omega_{\textrm{R}2}\\
    \end{pmatrix}.
    \label{eq:5x5matrix}
\end{equation}
\end{widetext} 
and in the dressed singlet-triplet basis $\{\ket{\text{S}(0,2)}, \ket{\text{T}_{+,\rho}},\ket{\text{S}(1,1)}, \ket{\text{T}_{0,\rho}} \ket{\text{T}_{-,\rho}} \}$
\begin{widetext}
\begin{equation}
H_{\rho,5\times5} =\frac{h}{2} 
    \begin{pmatrix}
     -2 \varepsilon && 0 && 2 t && 0 && 0\\
     0  && \Omega_{\textrm{R}1}+\Omega_{\textrm{R}2} && \frac{-\Delta\nu_1 + \Delta\nu_2}{\sqrt{2}} && \frac{\Delta\nu_1 + \Delta\nu_2}{\sqrt{2}} && 0\\
    2t_{\text{c}} && \frac{-\Delta\nu_1 + \Delta\nu_2}{\sqrt{2}} && 0 && \Omega_{\textrm{R}1}-\Omega_{\textrm{R}2} && \frac{\Delta\nu_1 - \Delta\nu_2}{\sqrt{2}}\\
     0 && \frac{\Delta\nu_1 + \Delta\nu_2}{\sqrt{2}} && \Omega_{\textrm{R}1}-\Omega_{\textrm{R}2} && 0  && \frac{\Delta\nu_1 + \Delta\nu_2}{\sqrt{2}}\\
     0 && 0 && \frac{\Delta\nu_1 - \Delta\nu_2}{\sqrt{2}} && \frac{\Delta\nu_1 + \Delta\nu_2}{\sqrt{2}} && -\Omega_{\textrm{R}1}-\Omega_{\textrm{R}2}\\
    \end{pmatrix}.
    \label{eq:5x5matrixST}
\end{equation}
\end{widetext}

\section{Initialization and Readout}
\begin{figure}[b]
    \centering
    \includegraphics[width=1\columnwidth]{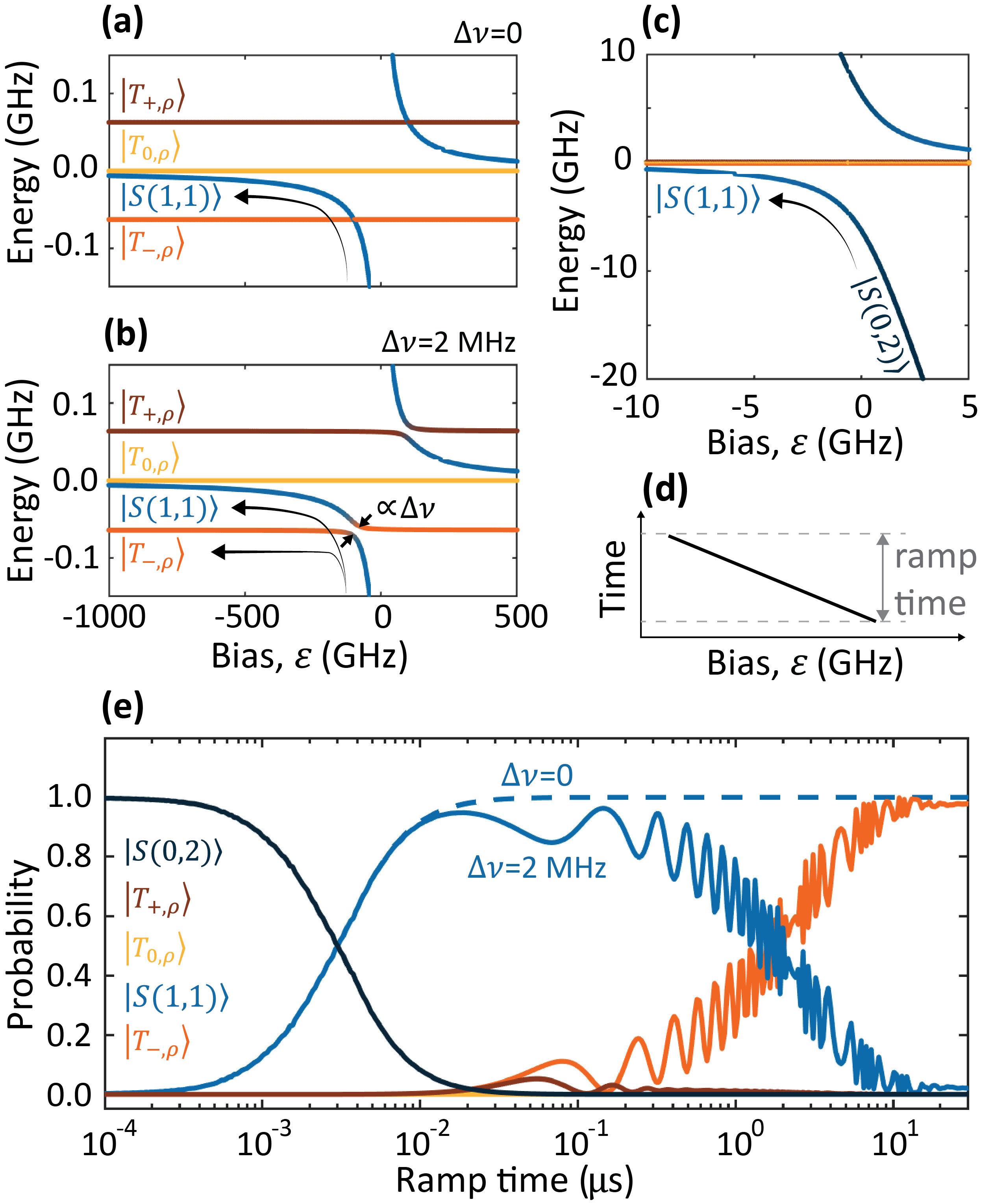}
    \caption{Initialization process in the dressed basis. Energy diagrams shown as a function of chemical potential bias $\varepsilon$ for the case of (a) no difference in Zeeman energies, $\Delta\nu_1 = \Delta\nu_2 = 0$ and (b) the driving frequency hitting the center of the Zeeman energy differences, $\Delta\nu_1 = -\Delta\nu_2$. The energy axis is extended in (c) to show $\ket{\text{S}(0,2)}$. The colors of the lines show the percentage of each state. The dark blue shows the $\ket{\text{S}(0,2)}$, blue is $\ket{\text{S}(1,1)}$, yellow is $\ket{\text{T}_{0,\rho}}$, brown is $\ket{\text{T}_{+,\rho}}$, and orange is $\ket{\text{T}_{-,\rho}}$. The black arrows show the path of initialization, either crossing the anticrossing [in both (a) and (b)] or avoiding it [only in (b)]. The initialization sequence is shown in (d) with time on the vertical axis, and detuning on the horizontal. This shows a positive $\varepsilon$, starting as $\ket{\text{S}(0,2)}$, followed by ramping $\varepsilon$ for a particular ramp time, finally measuring the state probability at a negative $\varepsilon$. This state probability against the ramp time is plotted for (e) $\Delta\nu_1 = \Delta\nu_2 = 0$ (dashed line) showing $\ket{\text{S}(1,1)}$ initialization and $\Delta\nu_1 = -\Delta\nu_2$ (solid line) showing $\ket{\text{T}_{-,\rho}}$ initialization. Values used: $t_{\text{c}}$ = 1 GHz, $\Omega_{\text{R1}}=\Omega_{\text{R2}}=10$ MHz and ramping range from $\varepsilon  = 50 \rightarrow 1500$ GHz.}
    \label{fig:anticrossing_comparison}
\end{figure}

The experimental process of preparing different two-qubit states is modelled following the dynamics governed by the Hamiltonian $H_{\rho,5\times5}$. Beginning in the $\ket{\text{S}(0,2)}$ ground state, $\varepsilon$ is adjusted to load an electron into the second dot to prepare the spatially separated spin states ($\ket{\text{T}_{+,\rho}}, \ket{\text{T}_{0,\rho}}, \ket{\text{T}_{-,\rho}}$ and $\ket{\text{S}(1,1)}$). The rate at which $\varepsilon$ changes determines which state is prepared; this is analogous to the initialization method in the rotating bare spin basis~\cite{Fogarty2018IntegratedReadout}. 

The behavior of each eigenenergy in the dressed five-level system is investigated as a function of $\varepsilon$. This is shown in Fig.~\ref{fig:anticrossing_comparison}(a,b,c), where the different colors represent the percentage of each state. For each initialization, the chemical potential bias $\varepsilon$ is ramped at a constant rate from a positive bias [which favors a (0,2) ground state] to a negative one, see Fig.~\ref{fig:anticrossing_comparison}(d). The values used in the simulation are $t_{\text{c}}$ = 1 GHz, $\Omega_{\text{R1}}=\Omega_{\text{R2}}=10$ MHz and ramping range from $\varepsilon  = 50 \rightarrow 1500$ GHz. To start with, the case where $\varepsilon$ is ramped to transfer $\ket{\text{S}(0,2)}$ into $\ket{\text{S}(1,1)}$ is considered. In an idealized case, one can set $\Delta\nu_1 = \Delta\nu_2 = 0$ so that there is no coupling term between the singlet and $\ket{\text{T}_{-,\rho}}$, demonstrated in Fig.~\ref{fig:anticrossing_comparison}(a) with an absence of an anticrossing between $\ket{\text{S}(1,1)}$ and $\ket{\text{T}_{-,\rho}}$ energy lines, allowing for a perfectly diabatic transition from $\ket{\text{S}(0,2)}$ to $\ket{\text{S}(1,1)}$ (shown by the black arrow). 

In this idealization, the transition into $\ket{\text{S}(1,1)}$ is dependent on how fast $\varepsilon$ changes due to the $\ket{\text{S}(0,2)}-\ket{\text{S}(1,1)}$ anticrossing, which can be controlled to be large by tuning $t_{\text{c}}$, as well as the difference in $\Omega_{\text{R1}}$ and $\Omega_{\text{R2}}$. Ideally, $\Omega_{\text{R1}}=\Omega_{\text{R2}}$ to avoid S-T mixing. The inverse of the tunnel coupling sets the time scale of the fastest allowable ramp, before a diabatic passage through the $(0,2)\rightarrow(1,1)$ transition occurs. As the ramp time is increased, the probability of preparing $\ket{\text{S}(1,1)}$ via an adiabatic crossing is increased. The probability of preparing each state at the end of the ramp sequence against the ramp time is plotted in Fig.~\ref{fig:anticrossing_comparison}(e) with dashed lines. Since $\Delta\nu_1 = \Delta\nu_2 = 0$, the only interacting states are the singlet states, so the figure shows the reducing probability of initializing $\ket{\text{S}(0,2)}$ as the ramp time increases, and the increasing probability of initializing $\ket{\text{S}(1,1)}$. A ramp time of 1 $\upmu$s is sufficient to initialize $\ket{\text{S}(1,1)}$ for the parameters used here.

Moving on to the initialization of $\ket{\text{T}_{-,\rho}}$, we look at the case where $\Delta\nu_1 = -\Delta\nu_2=2$ MHz and the other variables remain the same. An anticrossing between $\ket{\text{S}(1,1)}$ and $\ket{\text{T}_{-,\rho}}$ is now present, as shown in Fig.~\ref{fig:anticrossing_comparison}(b). As before, $\varepsilon$ is ramped from a positive energy to a negative one for different ramp times. The probability of preparing each state against the ramp time is plotted in Fig.~\ref{fig:anticrossing_comparison}(e) with solid lines. It is obvious that the introduced $\Delta\nu_1 = -\Delta\nu_2$ condition has allowed for the $\ket{\text{T}_{-,\rho}}$ to be initialized after approximately 110 $\upmu$s. For shorter ramp times, when the energy gap between the singlet and $\ket{\text{T}_{-,\rho}}$ is crossed diabatically, the initialised state develops components of both $\ket{\text{S}(1,1)}$ and $\ket{\text{T}_{0,\rho}}$ due to the coupling created by the difference in Larmor frequencies. When the ramp becomes fast enough ($1 \upmu$s), the $\ket{\text{S}(1,1)}$ initialisation is recovered. 

The importance of the lowest energy anticrossing becomes more apparent when the variability between the different qubit environments is regarded. The values of the $g$-factors are different in a pair quantum dots~\cite{tanttu_controlling_2019}, therefore the values of $\Delta\nu_1$ and $\Delta\nu_2$ will typically be different. Although the external magnetic field can be rotated to an angle that minimizes the difference between a single pair of $g$-factors~\cite{tanttu_controlling_2019}, for larger scale systems there will still be variability. This means that the scenario including the lowest energy anticrossing [Fig.~\ref{fig:anticrossing_comparison}(b)] is more realistic. 

Readout of a dressed qubit follows a similar method to initialization. Instead of ramping $\varepsilon$ from positive to negative, the reverse is implemented. The ramping is chosen at a particular rate so that it allows for $\ket{\text{S}(1,1)}$ to tunnel into $\ket{\text{S}(0,2)}$, but not the triplet states. This is the same singlet-triplet readout technique used for rotating bare spin qubits. Dressed parity readout is also achievable when considering dephasing in the system, so it follows similar dynamics to the rotating bare spin case~\cite{seedhouse_parity_2020}. 

\section{Single qubit gates} \label{sec:simgle_qubit_control}

\begin{figure}
    \centering
    \includegraphics[width=\columnwidth]{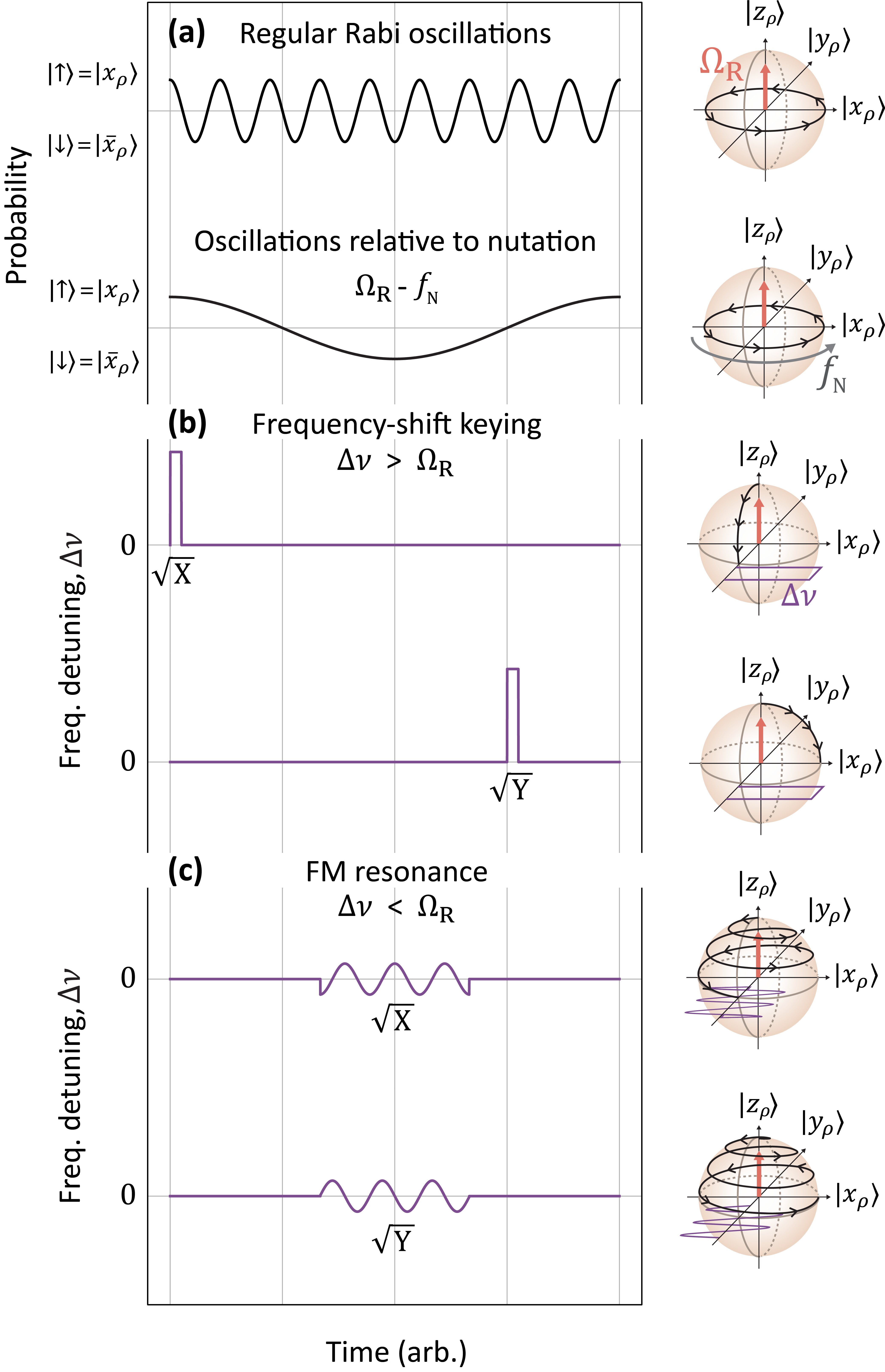}
    \caption{Bloch sphere representation of single qubit control of the dressed qubit. The precession about the $x-y$ plane according to (a) the Rabi frequency and the difference between the Rabi frequency and the nutating frequency $f_{\text{N}}$. (b) the frequency-shift keying method for single qubit control.The purple lines show the keying pulses of $\Delta\nu$ for $x$ and $y$ control. The Bloch spheres to the right of the plot show the $\sqrt{\rm X}$ and $\sqrt{\rm Y}$ rotations of the dressed qubit following the black line, where the initial state of the qubit is along $\ket{z_{\rho}}$. The pink arrow shows the direction of $\Omega_{\text{R}}$. The FM resonance method (c) in the rotating frame. The purple lines show $\sqrt{\rm X}$ and $\sqrt{\rm Y}$  gate pulses on $\Delta\nu$. The Bloch spheres are in the rotating frame and show the $\sqrt{\rm X}$ and $\sqrt{\rm Y}$  dressed qubit control where the initial state of the qubit is along $\ket{z_{\rho}}$.}
    \label{fig:single_qubit_control}
\end{figure}

For universal quantum computation, controllable rotations of the qubit system about two axes must be attainable. For the dressed qubit, single qubit gates can be achieved by pulsing the amplitude of $\Delta\nu$~\cite{laucht_dressed_2017}. In this section, we look at two different methods of pulsing $\Delta\nu$: frequency-shift keying and frequency modulation (FM) resonance. We begin by studying frequency-shift keying, considering the single qubit subspace in Eq.~(\ref{eq:single_dressed_qubit}). Frequency-shift keying is a frequency modulation scheme where changes to the frequency are discrete. In other words, $\Delta\nu$ is modulated with a square pulse. 

To see how two-axes control arises from frequency-shift keying, the qubit state should be described from a frame that is nutating at a frequency slightly higher or lower than the Rabi frequency. The Bloch spheres in Fig.~\ref{fig:single_qubit_control}(a) demonstrate this transformation; the top plot and Bloch sphere shows the Rabi frequency and the second plot and Bloch sphere shows the difference between the Rabi and frame nutation frequencies chosen. This transformation adds a time dependence to $\Delta\nu$, such that the square pulses can be timed to be at specific relative phases with respect to each other, leading to a $\sqrt{\rm X}$ or $\sqrt{\rm Y}$ gate. The gate pulses are shown by the purple lines in Fig.~\ref{fig:single_qubit_control}(b), as well as the Bloch spheres showing the dressed spin performing $\pi/2$ gates. Note that tracking phases and modulating synchronously the microwave and the qubit frequencies does not demand fast control since these modulations occur in the easily accessible radio-frequency range.

Returning to the rotating frame, we discuss FM resonance. In this case, $\Delta\nu$ is modulated using a sinusoidal shape. The frequency with which the modulation should take place is $\Omega_{\text{R}}$, causing a resonance with regard to the dressed qubit quantisation set by the driving field. With a sine modulation, rotations occur about the $x$ axis as shown in the top plot in Fig.~\ref{fig:single_qubit_control}(c). A phase can be added to the modulation so that the amplitude of $\Delta\nu$ follows a cosine wave, performing a rotation about $y$ as shown in the bottom plot in Fig.~\ref{fig:single_qubit_control}(c). One constraint of using this method is that when $\Delta\nu>\Omega_{\text{R}}$ the dressed qubit rotates at a faster rate than $\Omega_{\text{R}}$, breaking the rotating wave approximation and becoming sensitive to Bloch-Siegert shifts \cite{Bloch1940MagneticFields,Laucht2016BreakingSpin}. To avoid this, the amplitude of the modulation should satisfy $\Delta\nu\ll\Omega_{\text{R}}$ resulting in a pulse time that is long compared to $1/\Omega_{\text{R}}$. Otherwise, the control pulse needs to be especially engineered to account for the Bloch-Siegert oscillations.

Both control methods can lead to high fidelity single qubit gates. The choice of method depends on the controllability of $\Delta\nu$ for the particular qubit system, and how the amplitude of this control compares with the Rabi frequency.

Since we are in the dressed picture, there is a constant echoing of background noise due to the constant on-resonance field. In the regime where $\Delta\nu>\Omega_{\text{R}}$, the resonance effects become weak and noise reduction advantage of the dressed qubit is degraded. With this in mind, it is probable that the sinusoidal modulation of $\Delta\nu$ is the preferred method for single qubit control in most qubit systems.
 
\section{Two-qubit gates} \label{sec:two_qubit_control}

\begin{figure}
    \centering
    \includegraphics[width=1\columnwidth]{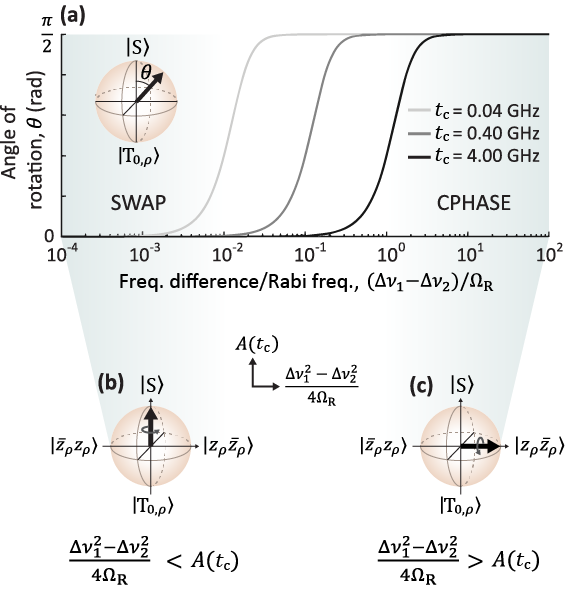}
    \caption{Two-qubit gate operations in the dressed basis. The comparison of eigenbasis from the Hamiltonian in Eq.~\ref{eq:reduced_ham} when the system parameters are adjusted. (a) The transition of the SWAP regime into the CPHASE regime determined by the eigenbasis along the vertical axis. The horizontal axis is the detuning frequency difference $(\Delta\nu_1-\Delta\nu_2)$ over the Rabi frequency of both qubits $\Omega_{\textrm{R}}$. Three curves are plotted, each for a different $t_{\text{c}}$, where $t_{\text{c}}=0.04, 0.4$ and $4$ GHz from left to right. A S-T Bloch sphere representation of the (b) SWAP gate and (c) the CPHASE gate, where the bold arrow shows the axis of rotation.}
    \label{fig:swap_cphase_bloch}
\end{figure}

It is important to understand the origin of two-qubit gates for universal quantum computing in the dressed picture. The intrinsic gates discussed here are the SWAP and CPHASE gate, both implemented by controllable exchange coupling between spins. In the rotating bare spin case, these gates have been discussed in detail~\cite{meunier_efficient_2011}, but there is no treatment of these gates for dressed qubits. Here, we focus on how the gradual change in the pulse ramp times leads to a shift from the SWAP to CPHASE gates by observing the evolution from Heisenberg exchange to Ising coupling. Following a similar analysis as in Ref.~\cite{meunier_efficient_2011} (see the Appendix~\ref{sec:A1} for details), the Hamiltonian of the dressed qubit in the $\{\ket{z_\rho \bar{z}_\rho},\ket{\bar{z}_\rho z_\rho}\}$ subspace is described in terms of a Schrieffer-Wolff transformation as
\begin{equation}
H_{\text{reduced}} =\frac{h}{2} 
    \begin{pmatrix}
     -A(t_{\text{c}})+\frac{\Delta\nu_1^2-\Delta\nu_2^2}{4\Omega} && A(t_{\text{c}}) \\
     A(t_{\text{c}}) &&  -A(t_{\text{c}})+\frac{-\Delta\nu_1^2+\Delta\nu_2^2}{4\Omega}\\
    \end{pmatrix},
    \label{eq:reduced_ham}
\end{equation}
\begin{equation}
    A(t_{\text{c}}) = \frac{t_{\text{c}}^2U}{U^2-\varepsilon^2}.
\end{equation}
Here we have introduced $U$ as the cost in energy for both electrons to be in the same dot (Hubbard $U$). The Schrieffer-Wolff approximation requires  $t_{\text{c}}<U\pm\varepsilon$, and we have assumed $\Omega_{\textrm{R}1} = \Omega_{\textrm{R}2} = \Omega_{\textrm{R}}$. 

Denoting the relative frequencies as $\Delta\nu_1-\Delta\nu_2$, a plot of the polar angle $\theta$ that defines the rotation axis on the singlet-triplet Bloch sphere as a function of $(\Delta\nu_1-\Delta\nu_2)/\Omega_{\textrm{R}}$ is shown in Fig.~\ref{fig:swap_cphase_bloch}(a). The difference in detuning frequencies over the Rabi frequency ratio is chosen because, in order for the dressed qubit to keep its echoing advantages, the detunings should not be larger than $\Omega_{\text{R}}$, meaning the ratio should be small (typically less than 0.1). The Figure shows that when the ratio $(\Delta\nu_1-\Delta\nu_2)/\Omega_{\textrm{R}}$ is small, $A(t_{\text{c}})$ is the dominating term in the Hamiltonian, resulting in the eigenbasis $\{\ket{\text{S}(1,1)},\ket{\text{T}_{0,\rho}}\}$, when $\theta=0$. It follows that oscillations between $\ket{z_\rho \bar{z}_\rho}$ and $\ket{\bar{z}_\rho z_\rho}$ occur, allowing for a SWAP gate. A singlet-triplet Bloch sphere in Fig.~\ref{fig:swap_cphase_bloch}(b) shows the rotation axis along $\ket{S}-\ket{\text{T}_{0,\rho}}$ in this case (precession). For larger $(\Delta\nu_1-\Delta\nu_2)/\Omega_{\textrm{R}}$, the eigenbasis shifts to $\{\ket{z_\rho \bar{z}_\rho},\ket{\bar{z}_\rho z_\rho}\}$ (nutation). For sufficiently large differences in qubit frequencies, $\theta=\pi/2$ giving a CPHASE gate, shown in Fig.~\ref{fig:swap_cphase_bloch}(c). It should be noted that when $t_{\text{c}}$ is increased or decreased, the region where the crossover between the two regimes occurs shifts to a higher or lower $(\Delta\nu_1-\Delta\nu_2)/\Omega_{\textrm{R}}$, respectively.  

The tunnel rates between spin qubits in quantum dots are typically tunable in the range 0.01-100 GHz, with larger values of $t_{\text{c}}$ shifting the crossover point in Fig.~\ref{fig:swap_cphase_bloch}(a) to a larger value of frequency differences. In other words, the transition from a SWAP gate to a CPHASE gate occurs at a larger $(\Delta\nu_1-\Delta\nu_2)/\Omega_{\text{R}}$ ratio. As mentioned before, the ratio $(\Delta\nu_1-\Delta\nu_2)/\Omega_{\text{R}}$ should not exceed 0.1 because the dressed qubit noise resilience is a consequence of $\Omega_{\text{R}}>\Delta\nu$. If a qubits is far detuned from the microwave frequency then the qubit is no longer dressed. As a result of the large tunnel coupling, the SWAP gate is the native two-qubit gate of the dressed qubit, since this gate lies in the region where $(\Delta\nu_1-\Delta\nu_2)/\Omega_{\text{R}}<0.1$ for tunnel couplings of the order of 1 GHz. 

\begin{figure}
    \centering
    \includegraphics[width=1\columnwidth]{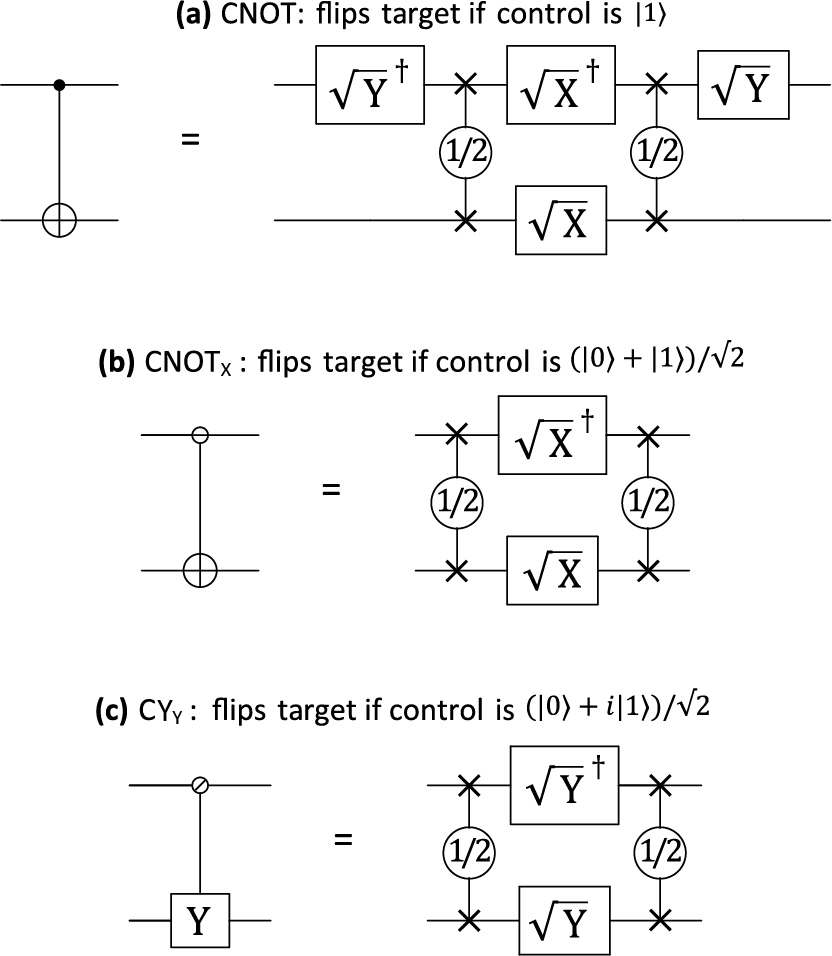}
    \caption{Conditional NOT gate decomposition where, if the control qubit is (a) $\ket{1}$, the target qubit is operated on by X. If the control is (b) $(\ket{0}+\ket{1})/\sqrt{2}$, the target qubit is rotated by X, and if the control qubit is (c) $(\ket{0}+i\ket{1})/\sqrt{2}$, the target qubit is rotated by Y. }
\label{fig:cnots}
\end{figure}

The particular choice of elementary two-qubit gate can also be customised depending on the particular application. A typical example would be the choice of gates for quantum error correction (QEC) within the surface code~\cite{fowler_surface_2012}. The stabiliser measurements in surface codes utilise the CNOT gate. Many qubit architectures have the native two-qubit gate as the CPHASE gate, making implementing CNOT gates natural. With SWAP being the native two-qubit gate for the dressed qubit, the CNOT gate has to be be composed using $\sqrt{\textrm{SWAP}}$ and single qubit operations, shown in Fig.~\ref{fig:cnots}(a). Other conditional gates include the CNOT$_{\text{X}}$ and CY$_{\text{Y}}$, which flip the target qubit conditional on the superposition states $(\ket{0}+\ket{1})/\sqrt{2}$ and $(\ket{0}+i\ket{1})/\sqrt{2}$, respectively. The notation for CNOT$_{\text{X}}$ is chosen because the gate is a conditional NOT operation on the logical basis, but the condition is that the control qubit state is along $x$; meanwhile, CY$_{\text{Y}}$ is chosen because the conditional rotation is, instead, about Y, and the condition is that the state is along $y$. These gates are shown in Fig.~\ref{fig:cnots}(b,c). It is clear from the figure that the CNOT$_{\text{X}}$ and CY$_{\text{Y}}$ have shorter circuit depth, which potentially leads to less errors in its implementation. A necessary condition for stabiliser measurements is that the syndromes commute with each other and that they measure orthogonal axes \cite{gottesman_1997_stabilizer,bennett_1996_mixed,knill_2000_theory}. Since the CNOT$_{\text{X}}$ and CY$_{\text{Y}}$ gates (when decomposed) are shorter, it is advantageous to have the $x$ and $y$ axes stabilized contructing them from CNOT$_{\text{X}}$ and CY$_{\text{Y}}$ gates.

\section{Scalability}

The main challenge for the scalability of a dressed qubit architecture is the ability to simultaneously drive all spins, which depends on the material-specific range of qubit frequency variability. We take the example of electron spins in silicon to discuss scalability.

In that case, the proposed method to shift $\Delta\nu$ is through the electrical control of the spin-orbit interactions of the electron. The two mechanisms responsible for spin-orbit interactions in silicon quantum dots are Rashba \cite{rashba_properties_1960} and Dresselhaus \cite{Dresselhaus_spin_1955} effects, originated by the reduction in crystal symmetry at the interface caused by the atomic scale disorder. Because of this, small variations in the structure (e.g. lattice imperfections, surface roughness) surrounding the quantum dots cause variations in the spin-orbit interactions, leading to differences in the $g$-factors. 

The external magnetic field angle can be tuned to align with the [1 0 0] direction of the silicon lattice and reduce the spin-orbit effects~\cite{tanttu_controlling_2019}. This minimises the Dresselhaus terms, leaving only a Rashba contribution. The magnetic field angle can also be tuned such that the difference in Dresselhaus and Rashba terms between dots cancel each other. However, this is only effective for two-dot systems since the relative strength of the two terms is different for each dot. Larger arrays of dots would mean that each dot sees a different environment such that cancelling the two effects in all dots simultaneously is impossible, hence a difference in $g$-factors is inevitable. For this proposal, differences in $g$-factors only become an issue if the spread of qubit frequencies caused by the $g$-factor variability is broader than the dressing field line width set by the Rabi frequency.

Note, however, that the electrical controllability of the Dresselhaus term is larger than the Rashba term~\cite{Ruskov2018ElectronDots}, which means that the logical gates can be implemented faster if the Dresselhaus spin-orbit coupling is not completely suppressed. The ideal angle of the magnetic field for a range of quantum dots to have good compromise between large controllability and small variability is left for future investigation, including an analysis of the microscopic sources of interface disorder.

The specific range of tolerable qubit frequencies can be improved if pulse-engineered methods are used to further decouple a qubit from noise and create an effective broader band for the driving field~\cite{Hansen_dyno_2021}.

\section{Summary}
In this manuscript we have proposed the use of a continuous on-resonance global field for a large spin qubit array, suitable for universal quantum computation. The implementation was compared to the rotating bare multi-spin qubit system with and without an off-resonance global field. From this it was clear that the dressed qubit system has the advantage of controlling a large number of qubits as well as being robust against noise.  With the logical states being encoded as dressed spins, the Pauli spin blockade remains active due to the rotational symmetry of singlet states, which allows for both initialization and readout. The protocol for performing single- and two-qubit operations has been shown, confirming that dressed spins are a suitable platform for scaling to large numbers of qubits. The effects of inhomogeneity between the $g$-factors of the qubits was discussed, concluding that the spread of $g$-factors should be within the linewidth set by the driving magnetic field. This constraint can be relaxed further by employing pulse engineering methods~\cite{Hansen_dyno_2021}.

Finally, the adoption of dressed qubits has a few implications for QEC. Firsly, the echoing properties of the always-on field decouple the qubit from time-correlated, non-Markovian noise. This is important because QEC codes deal better with Markovian noise. Another implication is that the native two-qubit gate is the SWAP gate for the dressed qubits; in the context of QEC, the two-qubit entangling gates have a significant impact. We have discussed the impact of the native SWAP gate, concluding that the CNOT$_{\rm X}$ and CY$_{\rm Y}$ should be utilized in quantum error correction with dressed spins to reduce circuit depth.

\section*{Acknowledgments}

We acknowledge support from the Australian Research Council (FL190100167 and CE170100012), the US Army Research Office (W911NF-17-1-0198), and the NSW Node of the Australian National Fabrication Facility. The views and conclusions contained in this document are those of the authors and should not be interpreted as representing the official policies, either expressed or implied, of the Army Research Office or the U.S. Government. The U.S. Government is authorized to reproduce and distribute reprints for Government purposes notwithstanding any copyright notation herein. I.H. and A.E.S. acknowledge support from Sydney Quantum Academy, NSW, Australia.

\appendix
\section{Schrieffer-Wolff transformation, finding the reduced Hamiltonian} \label{sec:A1}

Using the Schrieffer-Wolff transformation as demonstrated in \cite{Winkler_quasi_2003}, assuming $t_{\text{c}} < U \pm \varepsilon$, an effective Hamiltonian can be found describing the $\{ \ket{z_\rho \bar{z}_\rho},\ket{\bar{z}_\rho z_\rho} \}$ subspace. To begin with, the Hamiltonian describing the whole space $\{ \ket{z_\rho \bar{z}_\rho},\ket{\bar{z}_\rho z_\rho},\ket{\text{T}_{-,\rho}}, \ket{\text{T}_{+,\rho}},\ket{\text{S}(0,2)},\ket{\text{S}(2,0)}\}$
\begin{widetext}
\begin{equation}
H =\frac{h}{2} 
    \begin{pmatrix}
     0 && 0 && \Delta\nu_2 && \Delta\nu_1 && -\sqrt{2}t_{\text{c}} && -\sqrt{2}t_{\text{c}} \\
     0  && 0 && \Delta\nu_1 && \Delta\nu_2 && \sqrt{2}t_{\text{c}} && \sqrt{2}t_{\text{c}} \\
     \Delta\nu_2 && \Delta\nu_1 && -2\Omega_{\text{R}} && 0 && 0 && 0 \\
     \Delta\nu_1 && \Delta\nu_2 && 0 && 2\Omega_{\text{R}} && 0 && 0 \\
     -\sqrt{2}t_{\text{c}} && \sqrt{2}t_{\text{c}} && 0 && 0 && 2(U-\varepsilon) && 0 \\
     -\sqrt{2}t_{\text{c}} && \sqrt{2}t_{\text{c}} && 0 && 0 && 0 && 2(U+\varepsilon)
    \end{pmatrix}
\end{equation}
is split into $H = H_0 +H_1 + H_2$ such that $H_0$ includes the diagonal elements, $H_1$ the off-diagonal elements between the subspace of interest and the subspace to be removed, and $H_2$ the off-diagonal elements within these subspaces. 

\begin{equation}
H_0 =\frac{h}{2} 
    \begin{pmatrix}
     0 && 0 && 0 && 0 && 0 && 0 \\
     0  && 0 && 0 && 0 && 0 && 0 \\
     0 && 0 && -2\Omega_{\text{R}} && 0 && 0 && 0 \\
     0 && 0 && 0 && 2\Omega_{\text{R}} && 0 && 0 \\
     0 && 0 && 0 && 0 && 2(U-\varepsilon) && 0 \\
     0 && 0 && 0 && 0 && 0 && 2(U+\varepsilon)
    \end{pmatrix},
\end{equation}

\begin{equation}
H_1 =\frac{h}{2} 
    \begin{pmatrix}
     0 && 0 && \Delta\nu_2 && \Delta\nu_1 && -\sqrt{2}t_{\text{c}} && -\sqrt{2}t_{\text{c}} \\
     0  && 0 && \Delta\nu_1 && \Delta\nu_2 && \sqrt{2}t_{\text{c}} && \sqrt{2}t_{\text{c}} \\
     \Delta\nu_2 && \Delta\nu_1 && 0 && 0 && 0 && 0 \\
     \Delta\nu_1 && \Delta\nu_2 && 0 && 0 && 0 && 0 \\
     -\sqrt{2}t_{\text{c}} && \sqrt{2}t_{\text{c}} && 0 && 0 && 0 && 0 \\
     -\sqrt{2}t_{\text{c}} && \sqrt{2}t_{\text{c}} && 0 && 0 && 0 && 0
    \end{pmatrix},
\end{equation}

\begin{equation}
H_2 = 0.
\end{equation}
Setting $H'=H_1+H_2$ and denoting $H_0$ with $H^0$, the reduced Hamiltonian can be found from 
\begin{equation}
    H_{m m'} = H^0_{m m'}+H'_{m m'}+H'_{m l} H'_{l m'} \frac{1}{2} \sum_l \bigg( \frac{1}{E_m - E_l} + \frac{1}{E_m' - E_l} \bigg)
    \label{eq:SW}
\end{equation}
where $E$ are the eigenvalues of $H_0$, the indices $m,m'$ refer to the elements from the subspace $\{ \ket{\Psi_{\text{sym}}}, \ket{\text{T}_{-,\rho}} \}$, and $l,l'$ from the rest of the elements. 
Following equation~\ref{eq:SW}, the reduced Hamiltonian~(\ref{eq:reduced_ham}) is found,  
\begin{equation}
H_{\text{reduced}} =\frac{h}{2} 
    \begin{pmatrix}
     -\frac{t_{\text{c}}^2U}{U^2-\varepsilon^2}+\frac{\Delta\nu_1^2-\Delta\nu_2^2}{4\Omega} && \frac{t_{\text{c}}^2U}{U^2-\varepsilon^2} \\
    \frac{t_{\text{c}}^2U}{U^2-\varepsilon^2} &&  -\frac{t_{\text{c}}^2U}{U^2-\varepsilon^2}+\frac{-\Delta\nu_1^2+\Delta\nu_2^2}{4\Omega}\\
    \end{pmatrix}.
\end{equation}
\end{widetext}

\normalem
\bibliography{dressed}

\end{document}